\documentstyle[12pt,world_sci]{article}
\def\overlay#1#2{\ifmmode%
\setbox0=\hbox{$#1$}%
\setbox1=\hbox to\wd0{\hss$#2$\hss}\else%
\setbox0=\hbox{#1}%
\setbox1=\hbox to\wd0{\hss#2\hss}\fi%
 #1\hskip-\wd0\box1 }

\begin{document}
\hfill\vbox{\hbox{\bf NUHEP-TH-94-20}\hbox{Aug 1994}}\par

\title{\bf S-WAVE AND P-WAVE $B_c$ MESON PRODUCTION AT HADRON COLLIDERS BY
HEAVY QUARK FRAGMENTATION
}
\author{KINGMAN CHEUNG\footnote{To appear in proceedings of DPF'94 meeting,
 Albuquerque, NM (August 1994)} \\
{\em  Department of Physics, Northwestern University,
Evantson, Illinois  60208, U.S.A.\\}
\vspace{0.3cm}
and\\
\vspace*{0.3cm}
TZU CHIANG YUAN \\
{\em Davis Institute for High Energy Physics, Department of Physics, \\
University of California, Davis CA 95616, U.S.A.}
}

\maketitle
\setlength{\baselineskip}{2.6ex}

\begin{center}
\parbox{13.0cm}
{\begin{center} ABSTRACT \end{center}
{\small \hspace*{0.3cm}
We compute model-independently
the production rates and transverse momentum spectra for the $B_c$
mesons in various spin-orbital states ($n\,^1S_0$, $n\,^3S_1$, $n\,^1P_1$, and
$n\,^3P_J\,(J=0,1,2)$ ) at hadron colliders via the direct fragmentation of
the bottom antiquark and via the Altarelli-Parisi-induced gluon fragmentation.
Since all the radially and orbitally excited states below the $BD$
flavor threshold will decay, either electromagnetically, hadronically,
or a combination of both, into the pseudoscalar ground state $1\,^1S_0$,
they all contribute significantly to the inclusive $B_c$ meson production.
}}
\end{center}

The next and the last family of  $B$ mesons to be observed
will be the $B_c (\bar bc)$  made up of one charm quark and one bottom
antiquark.
Like the $J/\psi$ and $\Upsilon$ quarkonia,
dynamical properties of $B_c$ can be predicted reliably by using
perturbative QCD, in contrast to the heavy-light mesons.
In the limit $m_c/m_b \to 0$, the $B_c$ system
enables us to test the heavy quark symmetry and to understand
the next-to-leading terms in the heavy quark effective theory in the
applications to the heavy-light $B$ mesons.
In addition, the production rates for different
spin-orbital states also help us to understand the spin symmetry breaking
effects.   Phenomenologically,  $B_c$ mesons can be used to analyze the
mixing of the $B_s^0 - \overline {B_s^0}$ without ambiguity by tagging
the charge of the lepton in the decay $B_c^+ \to B_s^0 + \ell^+ \nu$ or
$B_c^- \to \overline{B_s^0} + \ell^- \bar \nu$

Calculations on the production of $B_c$ mesons at $e^+e^-$ colliders were
previously performed.   But the calculation for the production at
hadronic colliders is rather tedious
until Braaten and Yuan \cite{braaten} pointed out that the heavy quarkonium
production at the large transverse momentum region is dominated by heavy quark
fragmentation.
The fragmentation of a heavy quark into a heavy-heavy-quark bound state
essentially involves the creation of a heavy quark-antiquark pair, which tells
us that the process should be hard enough to be calculable in perturbative QCD
(PQCD).
Explicit calculation of the production rates and the transverse momentum
spectra was performed in Ref.~\cite{cheung}, which was based on the direct
$\bar b$ antiquark fragmentation functions $D_{\bar b \to B_c}(z)$ and
$D_{\bar b \to B_c^*}(z)$ calculated in Ref~\cite{bc}.  It was found
\cite{cheungyuan} that the Altarelli-Parisi-induced gluon fragmentation $g\to
B_c$ also contribute significantly to the total production.   Since the
$P$-wave fragmentation functions have just been completed \cite{yuan},
it is natural to include
all the $S$-wave and $P$-wave contributions to calculate the inclusive
production rate.

The direct $\bar b\to B_c$ fragmentation function and the induced $g\to B_c$
fragmentation function are coupled by the following Altarelli-Parisi
equations:
\begin{equation}
\label{Db}
\mu \frac{\partial}{\partial \mu} D_{\bar b\to B_c}(z,\mu) =
\int_z^1 \frac{dy}{y}
P_{\bar b\to \bar b}(z/y,\mu)\; D_{\bar b \to B_c}(y,\mu) +
\int_z^1 \frac{dy}{y} P_{\bar b\to g}(z/y,\mu)\; D_{g \to B_c}(y,\mu) \,,
\end{equation}
\begin{equation}
\label{Dg}
\mu \frac{\partial}{\partial \mu} D_{g\to B_c}(z,\mu) = \int_z^1 \frac{dy}{y}
P_{g \to \bar b}(z/y,\mu)\; D_{\bar b \to B_c}(y,\mu) +
\int_z^1 \frac{dy}{y} P_{g \to g}(z/y,\mu)\; D_{g \to B_c}(y,\mu) \,.
\end{equation}
where $P_{i\to j}$ can be approximated by the usual massless Altarelli-Parisi
splitting functions.   Similar equations can be written down for the
$^3S_1$, $^1P_1$, and $^3P_J\;(J=0,1,2)$ states.
The boundary conditions for the coupled equations are
$D_{g\to B_c}(z,\mu)=0$ for $\mu \le 2(m_b+m_c)$ and
$D_{\bar b \to B_c}(z,\mu_0=m_b+2m_c)$, which is the heavy quark fragmentation
function calculated to the order of $\alpha_s^2$ at the initial scale $\mu_0$
by PQCD.  Expressions for the initial fragmentation functions in $S$-wave
states can be found in Ref.~\cite{bc} and those for $P$-wave states can be
found in Ref~\cite{yuan}.

Numerically integrating the coupled equations with the above boundary
conditions, we obtain the direct $\bar b$ antiquark fragmentation functions
and the induced gluon fragmentation functions for the $S$-wave and $P$-wave
states at any arbitrary scale $\mu \ge \mu_0$.
The inputs to these fragmentation functions are the quark masses $m_b$ and
$m_c$ and the nonperturbative parameters associated with the wavefunction
of the bound state.  These nonperturbative parameters can be calculated within
the framework of the potential models \cite{quigg}.
For the two $S$-wave states there is only one nonperturbative parameter, which
is the radial wavefunction $R(0)$ at the origin.  The $P$-wave fragmentation
functions have two nonperturbative parameters associated with the
color-singlet and color-octet mechanisms.
Two of the $P$-wave states ($^1P_1$ and $^3P_1$)
mix to form two physical states because they have
the same quantum numbers.  The two physical states are denoted by
$|1+ \rangle$ and $|1+' \rangle$.  The mixing and further
details can be found in Ref~\cite{yuan}.

The calculation of $B_c$ meson production  is simplified by factorizing the
whole process into a short-distance process of producing the heavy quark and a
long-distance process, which is the fragmentation of the heavy quark into
the $B_c$ meson.  The differential cross-section for the $B_c$ meson in the
$^1S_0$ state is given by
\begin{eqnarray}
d\sigma\left(B_c   (p_T) \right)
&=& \sum_{ij} \int  f_{i/p}(x_1,\mu) f_{j/p}(x_2,\mu) \;  \left[
d \hat \sigma ( ij \to \bar b(p_T/z) X,\,\mu) \; D_{\bar b\to B_c}(z,\,\mu)
\right. \nonumber \\
&& \quad + \left.
d \hat \sigma ( ij \to g(p_T/z) X,\,\mu) \; D_{g\to B_c}(z,\,\mu) \right]\,,
\label{pt}
\end{eqnarray}
where $i,j$ denote all the possible initial partons.
The first term in the square bracket is the direct $\bar b$ fragmentation
contribution and the second term is the induced gluon fragmentation
contribution.
Similar expressions can be written down for the $^3S_1$ and the $P$-wave
states.   In the above equation, the factorization scale $\mu$ is chosen to be
of the order of $p_T/z$ to avoid large logarithms in $d \hat \sigma$,
while the large logarithms in $D(z)$ can be summed up by evolving the $D(z)$
according to the coupled equations in Eqs.~(\ref{Db}) and (\ref{Dg}).

\begin{figure}[t]
\vspace{4in}
\caption{\label{fig}
The differential cross sections $d\sigma/dp_T(B_c)$
versus the transverse momentum $p_T(B_c)$  of the $B_c$  meson for different
spin-orbital states at the Tevatron.
}
\end{figure}

In our calculation the factorization scale $\mu$ in Eq.~(\ref{pt}) is
chosen to be
\begin{equation}
\mu = \sqrt{ p_{T_{\bar b,g}}^2  + m_b^2}  \;,
\end{equation}
where $p_{T_{\bar b,g}}$ is the transverse momentum of the fragmentating
parton.  $p_{T_{\bar b,g}}$ is related to $p_T(B_c)$ by $p_{T_{\bar b,g}}=
p_T(B_c)/z$.
This scale is also used for the parton distributions and the running coupling
constant $\alpha_s$.  Explicitly, we used $m_b=4.9$ GeV, $m_c=1.5$ GeV, and
CTEQ2 \cite{cteq} for the parton distributions.  The $\alpha_s(Q)$ is
evaluated at 1-loop by evolving from the experimental value
$\alpha_s(m_Z)=0.118$ by $\alpha_s(Q)=\alpha_s(m_Z)/(1+((33-2n_f)/6\pi)
\alpha_s(m_Z)\log(Q/m_Z) )$, where $n_f$ is the number of active flavors at
the scale $Q$.
We included $gg\to b\bar b$, $g\bar b \to g \bar b$, and $q\bar q \to b\bar b$
as the hard subprocesses for the inclusive $\bar b$ production, and
$gg\to gg$, $gq(\bar q) \to gq(\bar q)$, and $q\bar q \to gg$ for the
inclusive $g$ production.
The fragmentation functions at the scale $\mu$ are obtained by solving
Eqs.~(\ref{Db}) and (\ref{Dg}) with the boundary conditions mentioned above.

The transverse momentum spectra for different spin-orbital states are shown in
Fig.~\ref{fig}, in which the contributions from the direct $\bar b$
fragmentation and the induced gluon fragmentation have been added,
 with the acceptance cuts
\begin{equation}
\label{cut}
p_T(B_c) > 10 \;{\rm GeV} \qquad {\rm and} \qquad |y(B_c)|<1
\end{equation}
on the $B_c$ mesons.   The curves in Fig.~\ref{fig} are for $n=1$ states.  The
radially excited $n=2$ states can be calculated similarly with the
corresponding nonperturbative parameters.    We present the integrated cross
sections for $n=1$ and $n=2$ states in Table~\ref{table}.

\begin{table}
\caption[]{\label{table}
The integrated cross sections in pb for the $B_c$ mesons in various
spin-orbital states, with the acceptance cuts in Eq.~(\ref{cut}), at the
Tevatron.  $|1+ \rangle$ and $|1+' \rangle$ are the two physical $P$-wave
states resulted from the mixing of the $^1P_1$ and $^3P_1$ states.
}
\medskip
\centering
\begin{tabular}{|c@{\extracolsep{0.5in}}|cc|}
\hline
           &   $n=1$     & $n=2$  \\
\hline
$^1S_0$    &   210       &  130   \\
$^3S_1$    &   350       &  210    \\
$^3P_0$    &   17        &  24     \\
$^3P_2$    &   38        &  54     \\
$|1+\rangle$     &  31   &  29     \\
$|1+'\rangle$    &  28   &  54     \\
\hline
\end{tabular}
\end{table}

Since the annihilation channel for the decay of the excited $B_c$ meson states
is highly suppressed relative to the electromagnetic and hadronic transitions,
all the excited states below the $BD$ threshold will decay into the ground
states by emitting photons or pions.   Thus, they all contribute to the
inclusive production.  Adding all the contributions shown in
Table~\ref{table}, we have a total cross section of 1.2 nb, which implies about
$1.2\times 10^5$ $B_c$ mesons for 100 pb$^{-1}$ at the Tevatron.
This number should
almost represent the total inclusive rate, except for a small contribution
{}from the $D$-wave states.

Thus, we have presented the so far most complete $B_c$ meson production via
heavy quark fragmentation, including $S$-wave and $P$-wave contributions,
at the Tevatron.
The $B_c$ meson can be detected via the decays into $J/\psi + X$, in which the
$J/\psi$ is fully reconstructed by the leptonic decay.  If $X$ is a charged
lepton, then the event has a very distinct signature of three charged
leptons coming out from the same displaced vertex.  If $X$ can be fully
reconstructed, together with the reconstructed $J/\psi$ the $B_c$ meson can be
fully reconstructed.

This work was supported by the U.~S. Department of Energy, Division of
High Energy Physics, under Grants DE-FG02-91-ER40684 and DE-FG03-91ER40674.

\bibliographystyle{unsrt}

\end{document}